# Electromagnetic Absorption in an Anisotropic Layered Superconductor

Mario Palumbo and M. J. Graf*

*Physikalisches Institut, Universität Bayreuth, D-95440 Bayreuth, Germany*
*\*Department of Physics & Astronomy, Northwestern University, Evanston, IL-60208, U.S.A.*

(July 19, 1995)

We calculate the infrared conductivity tensor of a layered superconductor considering two different order parameter symmetries: strongly anisotropic $s$-wave with line nodes, and pure $(d_{x^2-y^2})$ $d$-wave. The calculations are performed within the quasiclassical theory of superconductivity and include the effects of non-magnetic scattering processes. We discuss to what extent measurements of the optical absorption can be relied upon to distinguish between these two order parameter symmetries.

PACS numbers: 74.25.Nf, 74.25.Fy, 74.72.-h

*Introduction.* The symmetry of the superconducting order parameter in the high-$T_c$ cuprates is currently the source of considerable scientific debate. A number of experiments have suggested the presence of $d$-wave pairing (see for example 1, 2), while recent ARPES measurements[3] appear consistent with an anisotropic $s$-wave state. This is an important issue to resolve since the symmetry of the superconducting order parameter is a vital component in the understanding of the underlying pairing mechanism at work in the high-$T_c$ systems.

To date much theoretical effort has been directed towards understanding the effects of strongly anisotropic pairing on the thermodynamic and equilibrium properties of a superconducting system. This work includes studies of the behavior of the order parameter and excitation spectrum in the presence of impurities[5–7] and interfaces.[8,9] These considerations have revealed *qualitative* differences between the properties of anisotropic $s$-wave order parameters in comparison to $d$-wave systems, however there exists no unambiguous experimental verification as yet. In this paper we seek to go beyond the consideration of equilibrium properties by calculating the frequency-dependent current response of an anisotropic layered superconductor to an externally applied electromagnetic field. A study of the temperature dependent microwave conductivity has recently been put forth by Hirschfeld *et al.*[4] and Borkowski & Hirschfeld.[10] We demonstrate that under certain circumstances the dynamical properties of anisotropic superconductors are highly sensitive to the detailed structure of the order parameter, and thus may serve as a reliable probe for the order parameter anisotropy.

We perform our calculations within a microscopic model for layered superconductors which incorporates the effects of non-magnetic scattering processes. Considering two different order parameter symmetries (strongly anisotropic $s$-wave with lines of nodes, and pure $(d_{x^2-y^2})$ $d$-wave), we calculate both the in-plane and $c$-axis current response. We observe striking *qualitative* differences in the electrical current response of the different order parameter symmetries which are strongly correlated to the degree of (non-magnetic) scattering in the system. These differences are *not* due just to the differences in the excitation spectrum, but rather arise from the formation of a band of optically active Andreev bound states at frequencies below the gap edge.

*Microscopic Model.* We consider the microscopic model discussed in Refs. 11, 12 (the interlayer diffusion model), which is based on the quasiclassical theory of superconductivity. This model is characterized by an infinite periodic stack of *incoherently* coupled two-dimensional Fermi liquids. The in-plane transport is taken to be of the usual Fermi liquid type (i.e. mediated by charged quasiparticles propagating coherently with an in-plane velocity $v_f$). The interlayer transport, on the other hand, is diffusive in nature, originating from incoherent scattering processes. Interlayer scattering may take place through several different types of scattering processes such as electron-electron, electron-phonon, electron-impurity, etc. This model should be appropriate for systems whose $c$-axis transport is of the SIS (superconductor-insulator-superconductor-...) type. This type of behavior has recently been observed in a class of high-$T_c$ compounds,[13] suggesting that this model may be appropriate for certain high-$T_c$ materials.

A detailed description of the interlayer diffusion model, along with the derivation of both the in-plane and $c$-axis frequency-dependent conductivity, has been given elsewhere[11,12] and will not be repeated here. Instead, we present only a brief summary of the key features of the model along with a discussion of the necessary phenomenological parameters.

For simplicity we assume an isotropic cylindrical Fermi surface which we parameterize by the angle $\phi$. Within the quasiclassical formulation of superconductivity, our model is completely defined by specifying the form of the scattering self-energy, $\hat{\Sigma}_\ell$. This quantity is conveniently written as the sum of an in-plane and two interplane parts, $\hat{\Sigma}_\ell = \hat{\Sigma}_\ell^{\parallel} + \hat{\Sigma}_{\ell,\ell-1}^{\perp} + \hat{\Sigma}_{\ell,\ell+1}^{\perp}$, where $\ell$ is the layer index. In-plane scattering is taken to be isotropic so that the scattering self-energy can be written as $\hat{\Sigma}_\ell^{\parallel}(\epsilon,t) = c_i \hat{t}_\ell(\epsilon,t)$, where $\hat{t}_\ell$ is the single impurity $t$-matrix,

$$\hat{t}_\ell(\epsilon,t) = \hat{u}_0 + \hat{u}_0 \otimes \left[ N_f \oint \frac{d\phi}{2\pi} \hat{g}_\ell(\phi;\epsilon,t) \right] \otimes \hat{t}_\ell(\epsilon,t), \quad (1)$$

expressed in terms of the angular averaged single particle propagator, $\hat{g}_\ell$. Here $c_i$ is the effective concentration of scattering centers, $\hat{u}_0 = u_0 \hat{1}$ is an isotropic scattering potential, and $N_f$ is the total density of states (per spin) at the Fermi energy. Following Buchholtz and Zwicknagl,[14] we eliminate the parameters $c_i$ and $u_0$ in favor of an effective normal-state scattering rate, $1/\tau_\parallel$, and a normalized scattering cross-section, $\bar{\sigma}$. The normalized cross-section is a measure of the relative strength of the scattering and ranges from $\bar{\sigma} = 0$ for weak scattering (Born limit), to $\bar{\sigma} = 1$ for resonant scattering (unitarity limit).

We assume the $c$-axis coupling to be weak, and thus write the interlayer scattering self-energy in the Born approximation,

$$\hat{\Sigma}^\perp_{\ell,\ell\pm1}(\phi;\epsilon,t) = \hat{U}_{\ell,\ell\pm1}(t) \otimes \left[ \frac{\hbar}{2\pi} \oint \frac{d\phi'}{2\pi} \frac{1}{\tau_\perp(\phi,\phi')} \right.$$
$$\left. \times \hat{g}_{\ell\pm1}(\phi';\epsilon,t) \right] \otimes \hat{U}^\dagger_{\ell,\ell\pm1}(t). \quad (2)$$

The gauge operators, $\hat{U}_{\ell,\ell\pm1}$, are defined in terms of an averaged interlayer vector potential $A^z_{\ell,\ell\pm1}(t)$ by $\hat{U}_{\ell,\ell\pm1}(t) = \exp[-\frac{ied}{\hbar c}A^z_{\ell,\ell\pm1}(t)\hat{\tau}_3]$, where $d$ is the layer spacing. The effective interlayer scattering lifetime, $\tau_\perp(\phi,\phi')$, is taken to be anisotropic. We describe this anisotropy phenomenologically as $\tau_\perp(\phi,\phi') \propto \exp[-\gamma\cos(\phi-\phi')]$. The Fermi surface angles $\phi$ and $\phi'$ give the in-plane directions of the quasiparticle velocity before and after scattering to an adjacent layer. The parameter $\gamma$ specifies to what degree the scattered electrons "remember" their initial momentum. Isotropic scattering corresponds to $\gamma = 0$, while extreme forward scattering corresponds to $\gamma \to \infty$. Since we neglect coherent transport along the $c$-axis (i.e. we set the Fermi velocity along the $c$-axis to zero), the interlayer scattering self-energy is the *only* source of interlayer coupling in the model.

The interlayer diffusion model contains, as a minimal set, five material parameters: the transition temperature, $T_c$, the density of states at the Fermi level, $N_f$, the Fermi velocity, $v_f$, and the in-plane and inter-plane scattering lifetimes, $\tau_\parallel$ and $\tau_\perp$. All of these quantities can be deduced from normal-state measurements. In order to accommodate anisotropic pairing, we have also introduced a normalized in-plane scattering cross-section, $\bar{\sigma}$, and an interlayer scattering anisotropy parameter, $\gamma$.

The in-plane electrical current density is given in terms of the Keldysh component of the quasiparticle propagator, $\hat{g}^K_\ell$, by standard equations of Fermi liquid theory,[15]

$$\mathbf{j}_\ell(t) = eN_f \int \frac{d\epsilon}{4\pi i} \oint \frac{d\phi}{2\pi} \mathbf{v}_f(\phi) \text{Tr}\left\{\hat{\tau}_3 \hat{g}^K_\ell(\phi;\epsilon,t)\right\}, \quad (3)$$

where $\hat{\tau}_3$ is the third Pauli matrix. The microscopic expression for the interlayer current density was derived for isotropic interlayer scattering in Ref. 11, and is generalized below to anisotropic scattering,

$$j^z_{\ell,\ell+1}(t) = -\frac{eN_f d}{i\hbar} \int \frac{d\epsilon}{4\pi i} \oint \frac{d\phi}{2\pi} \text{Tr}\left\{\hat{\tau}_3 \left( \hat{\Sigma}^{\perp,R}_{\ell,\ell+1} \otimes \hat{g}^K_\ell \right.\right.$$
$$\left.\left. + \hat{\Sigma}^{\perp,K}_{\ell,\ell+1} \otimes \hat{g}^A_\ell - \hat{g}^R_\ell \otimes \hat{\Sigma}^{\perp,K}_{\ell,\ell+1} - \hat{g}^K_\ell \otimes \hat{\Sigma}^{\perp,A}_{\ell,\ell+1} \right) \right\}. \quad (4)$$

We compute the electrical conductivity by calculating $\mathbf{j}_\ell$ and $j^z_{\ell,\ell+1}$ in the presence of a weak electric field, and then reading off the appropriate coefficient. The procedure is rather involved and we thus refer the reader to Ref. 12 for the details of the calculation and the resulting expressions.

*Results.* We discuss the electromagnetic absorption in the superconducting state for each of the following order parameter models:

$$\Delta_{ASW}(\phi) = \Delta_0[1 + \cos(4\phi)]/2, \quad (5)$$
$$\Delta_{XSW}(\phi) = \Delta_0[1 + 3\cos(4\phi)]/4, \quad (6)$$
$$\Delta_{DW}(\phi) = \Delta_0 \cos(2\phi). \quad (7)$$

In our notation the subscripts $ASW$, $XSW$, and $DW$ denote anisotropic $s$-wave, extended $s$-wave, and $d$-wave, respectively. In terms of the irreducible representations of the $D_{4h}$ (tetragonal) group, $\Delta_{ASW}(\phi)$ and $\Delta_{XSW}(\phi)$ transform like the $A_{1g}$ (identity) representation, while $\Delta_{DW}(\phi)$ transforms like the $B_{1g}$ ($d_{x^2-y^2}$) representation. All three order parameters possess nodes on the Fermi surface, but only $\Delta_{XSW}(\phi)$ and $\Delta_{DW}(\phi)$ change sign.

We calculate the supercurrents and associated conductivities by numerically solving the quasiclassical transport equations.[11,12] These solutions must be carried out self-consistently for the order parameter amplitude, $\Delta_0$, the scattering self-energy, $\hat{\Sigma}_\ell$, and the quasiclassical propagator, $\hat{g}_\ell$. The conductivity can then be calculated from the analytic expressions derived in Ref. 12 by carrying out the necessary integrations.

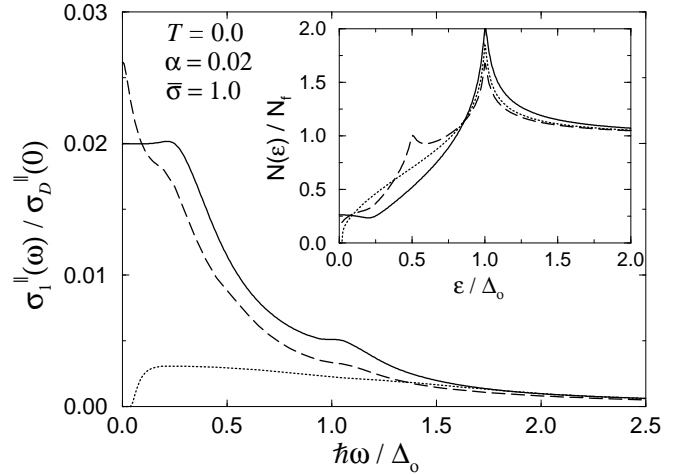

Figure 1. The in-plane infrared conductivity in the unitarity limit ($\bar{\sigma}=1$) for the $ASW$ (dotted line), $XSW$ (dashed line), and $DW$ (solid line) models in units of the d.c. Drude conductivity, $\sigma^\parallel_D(0) = e^2 N_f v_f^2 \tau_\parallel$. The inset shows the corresponding excitation spectra.

In figure 1 we show the real part of the in-plane frequency-dependent conductivity for the three order parameter models presented above, together with the corresponding excitation spectra (inset). These data are for a fairly clean system with a dimensionless scattering rate $\alpha \equiv \hbar/(2\pi\tau_\parallel k_B T_{c0}) = 0.02$, where $T_{c0}$ is the transition temperature in the absence of scattering. We consider here resonant scattering ($\bar{\sigma} = 1$), since it displays the most striking features, and we take the temperature equal to zero. As was pointed out previously,[5,7] even a small amount of scattering opens a gap in the excitation spectrum of the $ASW$ model, while the other two models display a finite density of states even at zero energy. The low energy enhancement in the density of states of the $XSW$ and $DW$ models can be interpreted as a band of optically active Andreev bound states.[6,12] These qualitative features are also represented in the absorption spectrum. The $ASW$ order parameter has a finite gap for any nonzero lifetime, $\tau_\parallel$, which results in a vanishing absorption below a critical frequency $\omega < \omega_{cr} \sim 1/\tau_\parallel$. The $XSW$ and $DW$ order parameters still possess nodes, however, and exhibit a significantly increased absorption for $\hbar\omega \lesssim \Delta_0$. The enhanced absorption at low $\omega$ has a width of the order of the cross-over energy $\epsilon^* \sim \sqrt{\hbar\Delta_0/4\tau_\parallel}$, and comes from transitions within the bound band (resonant scattering), while the small feature at $\hbar\omega \approx \Delta_0$ is a result of transitions from the bound band to the gap edge. Figure 1 demonstrates that the low-frequency behavior of the conductivity is highly sensitive to sign changes in the order parameter, even though the density of states displays rather slight differences.

Surprisingly, the $\omega \to 0$ absorption in the $XSW$ model is actually larger than in the $DW$ model. This difference can be *quantitatively* accounted for in terms of phase-space arguments. At $T=0$ one can show that $\sigma_1^\parallel(\omega \to 0)$ is just proportional to (the total number of nodes in $\Delta(\phi)$) $\times$ (the slope of $\Delta(\phi)$ at the nodes)$^{-1}$. Hence, for our order parameter models, $\sigma_{DW}^\parallel(\omega \to 0) \sim (4) \times (2)^{-1}$ which is less than $\sigma_{XSW}^\parallel(\omega \to 0) \sim (8) \times (2\sqrt{2})^{-1}$ by a factor of $1/\sqrt{2}$ (plus terms of the order $\sim \hbar/[\tau_\parallel \Delta_0]$). This estimate is in quantitative agreement with the numerical result in figure 1. Our analysis can be set on a more rigorous footing by noting that for rather clean systems ($\epsilon^{*2} \ll \Delta_0^2$) at $T = 0$, the low frequency limit of the in-plane conductivity can be written approximately as

$$\sigma_1^\parallel(\omega \to 0) \simeq e^2 N_f v_f^2 \hbar \oint \frac{d\phi}{2\pi} \frac{\epsilon^{*2}\cos^2(\phi)}{[\tilde{\Delta}_r^2(\phi) + \epsilon^{*2}]^{3/2}}, \quad (8)$$

where $\tilde{\Delta}_r(\phi)$ is the real part of the scattering-renormalized order parameter at $\epsilon=0$. We can represent $\tilde{\Delta}_r(\phi)$ in a very general way by

$$\tilde{\Delta}_r(\phi) = \Delta_0 \left[1 - \beta + \beta \cos(2n\phi) + \Delta_1\right], \quad (9)$$

where $\Delta_1 \sim 1/\tau_\parallel$ is the real part of the off-diagonal contribution to the scattering self-energy. Note that all three models being considered here may be represented in this way. If $\tilde{\Delta}_r(\phi)$ possesses nodes, then the largest contribution to the integral in equation (8) comes from the regions where $\tilde{\Delta}_r(\phi) \approx 0$. In this case we obtain the simple result:

$$\sigma_1^\parallel(\omega \to 0) \simeq \frac{e^2 N_f v_f^2 \hbar}{\pi \Delta_0} \frac{1}{\sqrt{\beta^2 - (1 - \beta + \Delta_1)^2}}, \quad (10)$$

which is valid as long as $min[\tilde{\Delta}_r(\phi)] \lesssim -\epsilon^*$. It is interesting to note that our result does not depend on the value of the "symmetry parameter" $n$ in equation (9).

Equation (10) implies that the quantity $\Delta_0 \sigma_1^\parallel(\omega \to 0)$ is relatively independent of both the scattering lifetime, $\tau_\parallel$, and cross-section, $\bar{\sigma}$; these quantities only enter indirectly through the scattering self-energy piece, $\Delta_1$. In fact, for the $DW$ model ($n = 1, \beta = 1$), the off-diagonal scattering self-energy vanishes ($\Delta_1 = 0$) so that $\Delta_0 \sigma_1^\parallel(\omega \to 0) \simeq e^2 N_f v_f^2 \hbar/\pi$, which is completely independent of the degree of scattering.[16] Figure 2 shows a series of absorption spectra for an $XSW$ order parameter with a dimensionless scattering rate $\alpha = 0.1$ for several different values of $\bar{\sigma}$. Note that the $\omega \to 0$ limits for the conductivity all lie within $\sim 10\%$ of each other while the zero-energy values of the corresponding excitation spectra are very different. In the case of weak scattering, the limiting regime is only realized at very low frequencies, however this region attains an appreciable width for larger cross-section values. It is important to point out that equation (10) *quantitatively* accounts for the $\omega \to 0$ limits in figure 2, and thus allows one, in principle, to obtain an estimate for the gap-anisotropy parameter $\beta$ from a knowledge of the low-frequency absorption spectrum.

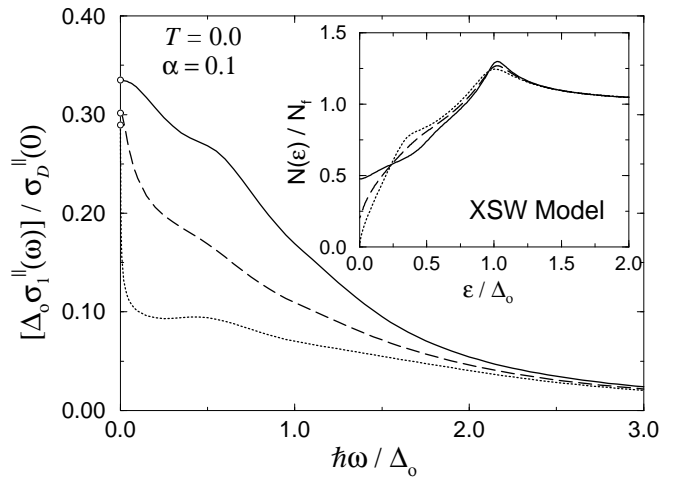

Figure 2. The in-plane conductivity in the $XSW$ model, for $\bar{\sigma} = 0.0$ (dotted line), $\bar{\sigma} = 0.5$ (dashed line), and $\bar{\sigma} = 1.0$ (solid line) in units of the d.c. Drude conductivity, $\sigma_D^\parallel(0) = e^2 N_f v_f^2 \tau_\parallel$. The $\omega \to 0$ limits are indicated by hollow circles. The inset shows the corresponding excitation spectra.

A similar analysis of the low-frequency limit for the c-axis conductivity shows that this frequency range is dominated by the in-plane scattering properties. For brevity we restrict our analysis to the diffuse transmission limit ($\gamma = 0$) and the specular transmission limit ($\gamma \to \infty$). Any finite $\gamma$-value will only change the absolute value of the result but not its physical behavior. For weakly coupled layers at zero-temperature, the c-axis conductivity has the general form:

$$\sigma_1^\perp(\omega \to 0) \simeq \frac{2e^2 d^2}{N_f} \oint \frac{d\phi}{2\pi} \oint \frac{d\phi'}{2\pi} \frac{N(\phi; 0) N(\phi'; 0)}{\tau_\perp(\phi, \phi')}, \quad (11)$$

where $N(\phi; 0)$ is the angle resolved density of states at $\epsilon = 0$, and $\phi$ and $\phi'$ refer to Fermi surface positions in two adjacent planes. In the diffuse transmission limit, $\tau_\perp^{-1}(\phi, \phi') \equiv \tau_\perp^{-1}$ and we obtain the simple result

$$\sigma_1^\perp(\omega \to 0) \simeq \frac{2e^2 d^2}{N_f \tau_\perp} \left\langle N(\phi;0) \right\rangle_\phi^2, \quad (12)$$

where $\langle \cdots \rangle_\phi$ denotes a Fermi surface average. This is what one expects for incoherent quasiparticle tunneling between two identical superconductors. Note that the zero-energy density of states, $\langle N(\phi;0)\rangle_\phi$, depends in a complicated way on the in-plane scattering parameters $\tau_\parallel$ and $\bar\sigma$. In the specular transmission limit, $\tau_\perp^{-1}(\phi,\phi') \equiv 2\pi\tau_\perp^{-1}\delta(\phi-\phi')$ and equation (11) becomes:

$$\sigma_1^\perp(\omega \to 0) \simeq \frac{2e^2 d^2}{N_f \tau_\perp} \left\langle N^2(\phi;0) \right\rangle_\phi. \quad (13)$$

Following the same analysis as in the case of the in-plane conductivity, we find for specular transmission the result:

$$\sigma_1^\perp(\omega \to 0) \simeq \frac{2e^2 N_f d^2}{\tau_\perp \Delta_0} \frac{\epsilon^*}{\sqrt{\beta^2 - (1-\beta+\Delta_1)^2}}, \quad (14)$$

which depends explicitly on the inter-plane scattering lifetime, $\tau_\perp$, and implicitly on the in-plane parameters $\tau_\parallel$ and $\bar\sigma$ through $\epsilon^*$ and $\Delta_1$. Equations (12) and (14) show that no universal behavior is expected in the inter-plane transport. In fact, the $c$-axis infrared absorption spectrum is very nearly a direct map of the density of states in the layers (for $\hbar\omega \lesssim \Delta_0$). Again one finds that the $ASW$ model opens a gap for any finite in-plane lifetime (always assuming that $\tau_\perp \gg \tau_\parallel$). Note that in general the cross-over energy $\epsilon^*$ is quite different for different anisotropic pairing states as well as for weak and strong scattering, providing a way to distinguish between these various scenarios.

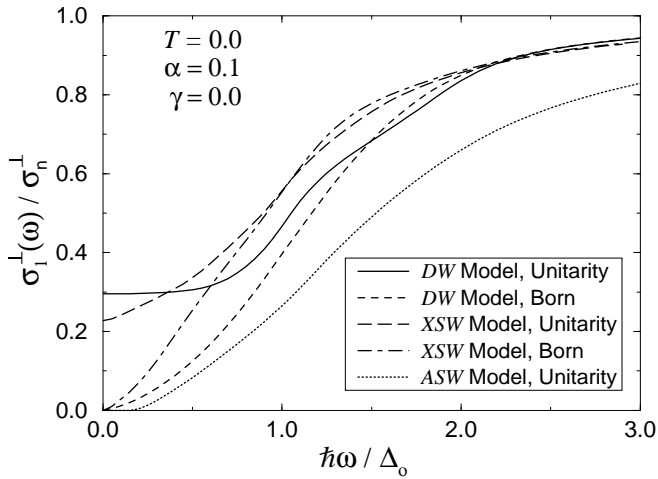

Figure 3. The $c$-axis infrared conductivity for our three order parameter models in units of the normal state value, $\sigma_n^\perp = 2e^2 N_f d^2/\tau_\perp$, for diffuse transmission.

In figure 3 we show the $c$-axis conductivity in the diffuse transmission limit for the three different order parameter models. For the $XSW$ and $DW$ models we have plotted results for both the Born and unitarity limits (the differences are very slight for the $ASW$ model). Note that the $XSW$ model no longer has a larger $\omega \to 0$ limit than the $DW$ case, nor do any of the models obey a universal low frequency limit. This is the behavior expected from equation (12), since the zero-energy density of states in the $XSW$ model is lower than that of the $DW$ model for this choice of parameters. The situation is not significantly altered in the case of specular transmission, except for slight quantitative differences. We note, however, that the value of $\gamma$ has a much more profound effect on the coherent $c$-axis transport (i.e. Josephson tunneling). In fact, one finds that a finite $\gamma$ is necessary for the $DW$ model to exhibit a finite Josephson effect.[17]

*Conclusion.* We have shown that anisotropic superconductors with lines of nodes exhibit, at $T \to 0$ and $\omega \to 0$, a strongly enhanced infrared absorption, and we have derived explicit expressions for its magnitude in terms of the order parameter anisotropy. The in-plane conductivity becomes universal for a pure $d$-wave order parameter,[16] while an extended $s$-wave pairing state displays a nearly universal behavior in the clean limit. Conversely, the $c$-axis conductivity (for diffuse interlayer coupling) is *not* universal, but rather resembles the density of states in the layers. Our results imply that one could, in principle, quantitatively ascertain the order parameter anisotropy through either a study of the scaling of the $\omega \to 0$ absorption with impurity concentration, or by searching for an impurity induced gap.

*Acknowledgments.* The authors thank D. Rainer and J.A. Sauls for many valuable discussions. The research of M.P. was supported by the Alexander von Humboldt-Stiftung, and that of M.J.G. was supported by the NSF (DMR 91-20000) through the Science and Technology Center for Superconductivity.